\documentclass{article}
\usepackage[T1]{fontenc} 
\usepackage[utf8]{inputenc} 

\usepackage{ismir,amsmath,cite,url}
\usepackage{graphicx}
\usepackage{multirow}
\usepackage{booktabs}
\usepackage{subcaption}
\usepackage{caption}

\usepackage{color}

\def\lbd{}	

\title{Decoding and Visualising Intended Emotion in an Expressive Piano Performance}

\multauthor
{Shreyan Chowdhury$^1$ \hspace{0.5cm} Gerhard Widmer$^{1,2}$} { 
$^1$Institute of Computational Perception, Johannes Kepler University Linz, Austria\\
$^2$LIT AI Lab, Linz Institute of Technology, Austria\\
{\tt\small {firstname.lastname}@jku.at}
}
\def\authorname{Shreyan Chowdhury, Gerhard Widmer}

\def\authorname{S. Chowdhury, and G. Widmer}

\begin{document}

\maketitle
\begin{abstract}
Expert musicians can mould a musical piece to convey specific emotions that they intend to communicate. In this paper, we place a \textit{mid-level} features based music emotion model in this performer-to-listener communication scenario, and demonstrate via a small visualisation music emotion decoding in real time. We also extend the existing set of mid-level features using analogues of \textit{perceptual speed} and \textit{perceived dynamics}.
\end{abstract}
\vspace{-2mm}
\section{Introduction}\label{sec:introduction}

Music emotion recognition aims at identifying and recognising emotional content in music using computer systems and typically considers \textit{perceived} emotion, which is the emotion that a human listener may recognise when listening to a song (that may be different from what the composer attempted to express and what the listener feels in response to it). However, in scenarios where the \textit{intended} emotion (emotion that the composer or performer aims to transmit)
is available (for instance in the experiments by \cite{Gabrielsson1996Emotional}), one could view emotion recognition models in a different light. Instead of comparing model predictions with listener ratings of perceived emotion, they could be compared directly to the intended emotion.

In the present paper, we are interested in placing a music emotion recognition model based on \textit{mid-level features} in a similar scenario -- decoding music emotion in real-time vis-à-vis a performer's intended emotions. (We have shown in previous research \cite{chowdhury2021perceived} that mid-level perceptual features, such as rhythmic complexity, or perceived major/minor harmonic character, are effective in modelling music emotion.) Also, in addition to the original seven mid-level features from \cite{Aljanaki2018Midlevel} and \cite{chowdhury2021perceived}, we investigate approximations of two additional mid-level features -- \textit{perceptual speed} and \textit{perceived dynamics}.

In Section \ref{sec:features}, we describe the original seven mid-level features and the two additional ones. In Section~\ref{sec:predict}, we compare emotion modelling using the original (7)-mid-level and the augmented (9)-mid-level feature sets and find that the two additional features improve emotion modelling substantially. 
Finally, in Section~\ref{sec:thevis}, we demonstrate how the new (9)-mid-level feature model can effectively decode, in real time, the specific emotions that a musician (multiple Grammy Award winner Jacob Collier) is explicitly trying to communicate in a spontaneous ``emotion communication experiment" that he shared with the world on YouTube \cite{jacob}.

\vspace{-4mm}
\section{The Features}\label{sec:features}


In previous research \cite{chowdhury2021perceived}, we have shown that training an emotion model based on mid-level perceptual features improves the model's accuracy and robustness, compared to models based on traditional low-level features, or to models that predict emotion in an end-to-end fashion directly from spectrograms. In this paper, we use the mid-level features based approach, however we investigate two additional features as well, which were absent in previous mid-level feature sets -- \textit{perceptual speed} and \textit{perceived dynamics}. 

\vspace{-2mm}
\subsection{Mid-level Features}
Mid-level features are musical qualities that are supposed to be
intuitively recognisable by most listeners, without requiring music-theoretic knowledge. In \cite{Chowdhury2019} and \cite{chowdhury2021perceived}, we used mid-level feature predictors trained on human-annotated data from \cite{Aljanaki2018Midlevel} so that our emotion recognition model using these would conform to how humans seem to perceive
mid-level qualities, and we showed that emotion recognition based on these features results both in good recognition accuracy, and in musically interpretable models. The set of seven mid-level features used in these works are: \textit{melodiousness}, \textit{articulation}, \textit{rhythm complexity}, \textit{rhythm stability}, \textit{dissonance}, \textit{tonal stability}, and \textit{minorness} (or \textit{mode}).


\vspace{-1mm}
\subsection{Perceptual Speed and Dynamics}\label{sec:speed_and_dyn}
While music emotion is modelled sufficiently well using the original seven mid-level features
defined in \cite{Aljanaki2018Midlevel},
two important features for emotional expression are conspicuously missing from this set: \textit{perceptual speed} and \textit{dynamics} \cite{bresin2011emotion, eerola2013emotional}. Previous studies \cite{eerola2013emotional} have indicated that musical cues such as attack rate and playing dynamics play a significant role in musical emotion expression. Our hypothesis is that augmenting the mid-level feature space with features analogous to perceptual speed and dynamics should improve emotion modelling significantly. Lacking appropriate human-annotated datasets, we model these features in a more direct way based on our musical intuition rather than on empirical listener perception data.

\subsubsection{Perceptual Speed}
Perceptual speed indicates the general speed of the music disregarding any deeper analysis such as the tempo, and is easy for both musicians and non-musicians to relate to \cite{friberg2014using}. While tempo is typically computed as the occurrence rate of the most prominent metrical level (beat), perceptual speed is influenced by lower level or higher level metrical levels as well. We approximate perceptual speed using \textit{onset density}, following observations from \cite{madison2010ratings, elowsson2013modelling}. Onset density (or event density) refers to the number of musical onsets (the beginnings of musical notes or other sonic events) per unit time. 

While several signal processing based onset density detectors are available in practice (for instance, the SuperFlux algorithm \cite{bock2013maximum} implemented in the \texttt{madmom} library \cite{madmom}), we take a different approach for our case. Since here we are aiming to only consider solo piano performances, we use an automatic polyphonic piano note transcription algorithm (\texttt{RNNPianoNoteProcessor} from \texttt{madmom} \cite{madmom}) to transcribe the audio and extract the piano notes, from which we then infer the onsets (the starting positions of the transcribed notes are taken as the onsets). We observed that this procedure gives us a better estimate of onsets for solo piano music across different recordings.

\subsubsection{Perceived Dynamics}
Perceived dynamics indicates the played dynamic level disregarding listening volume (presumably related to the estimated effort of the player) \cite{friberg2014using}. In our case of solo piano music, we find that the RMS (Root-Mean-Squared) amplitude of the audio signal is a good approximation to performed dynamics. We use Librosa's RMS function \cite{mcfee2015librosa} to compute this feature. For an input audio signal $x$, the RMS amplitude is given as $\text{RMS}_k = \sqrt{\text{mean}(w_\tau(x)_k^2)}$, $k = 1 \dots N$, where $w_\tau(\cdot)_k$ is a rectangular windowing function which partitions the input sequence into frames and returns the $k$-th frame of length $\tau$, and $N$ is the total number of frames.

    
    

    
    

\vspace{-3mm}
\section{Predicting Emotion}\label{sec:predict}
In a previous publication \cite{chowdhury2021perceived}, we reported the predictive performance for music emotion using different feature sets (including the original (7)-mid-level features set) on a dataset comprising of recordings of Bach's Well-Tempered Clavier (WTC) Book 1. Here, again, we use the WTC dataset\footnote{The WTC dataset contains recordings of 48 pieces of Bach's Well-Tempered Clavier Book~1 performed by 6 different pianists and multiple listener ratings of arousal and valence for each recording.} to evaluate the predictive performance of the original and augmented mid-level feature sets. We perform a regression-based analysis by fitting the different feature sets on the arousal/valence ratings. 

First, we predict the original (7)-mid-level features for the 288 recordings using a pre-trained mid-level feature model (trained on the Mid-level Features Dataset). The architecture of this model is a modified receptive-field regularised ResNet (see Koutini et al. \cite{koutini2019receptive}) domain-adapted for solo piano audio as described in Chowdhury and Widmer \cite{chowdhury2021towards}. Since in Section 4, our aim is going to be to predict emotions continuously using a sliding window, we want the input window size to be as small as possible without compromising model performance. Through experiments detailed in Chowdhury \cite{chowdhury2022thesis}, we find that an input window length of 5 seconds is optimal for our purpose. 

We then compute the mean onset densities and mean RMS amplitudes for each of the recordings using the approximations mentioned in Section~\ref{sec:speed_and_dyn}, giving us the (9)-mid-level feature set for the Bach dataset. A multiple linear regression model with nine inputs and two outputs (for arousal and valence) is then fitted on the data, and we report the adjusted $R^2$-score. This is compared to the case where only the original seven (7)-mid-level features are used, and where only the two newly added features are used. The results are tabulated in Table~\ref{tbl:perf}.

\begin{figure}
\footnotesize
\captionsetup{font=small}
\centering
    \vspace{-4mm}
    \begin{subfigure}{.5\textwidth}
        \caption{T-values for Arousal}
        \includegraphics[trim={0 0 0 0}, clip, width=\columnwidth]{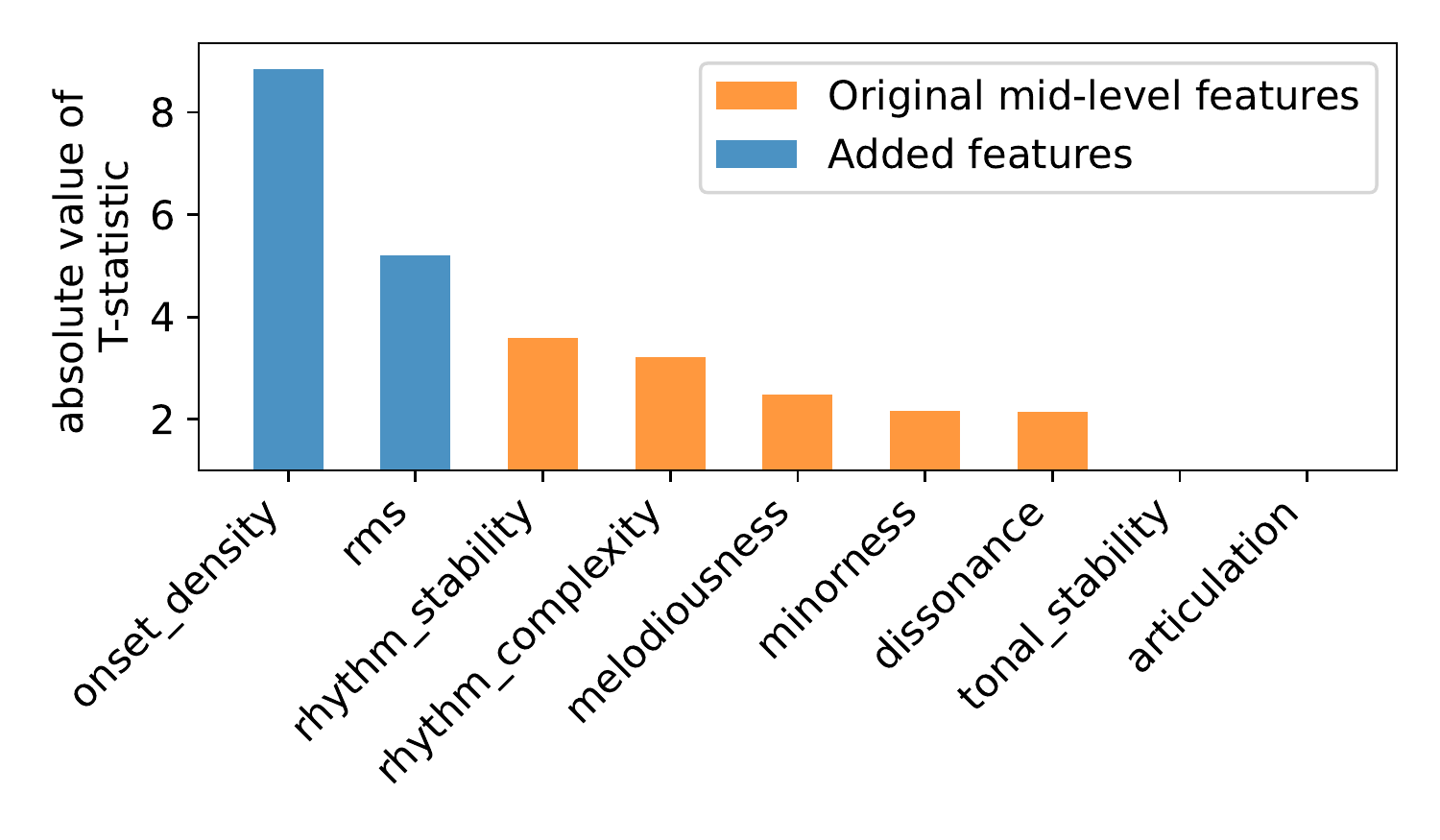}
        \label{ch_disentangle:fig:tvals_arousal}        
    \end{subfigure}\\
    \vspace{-4mm}
    \begin{subfigure}{.5\textwidth}
        \caption{T-values for Valence}
        \includegraphics[trim={0 0 0 0}, clip, width=\columnwidth]{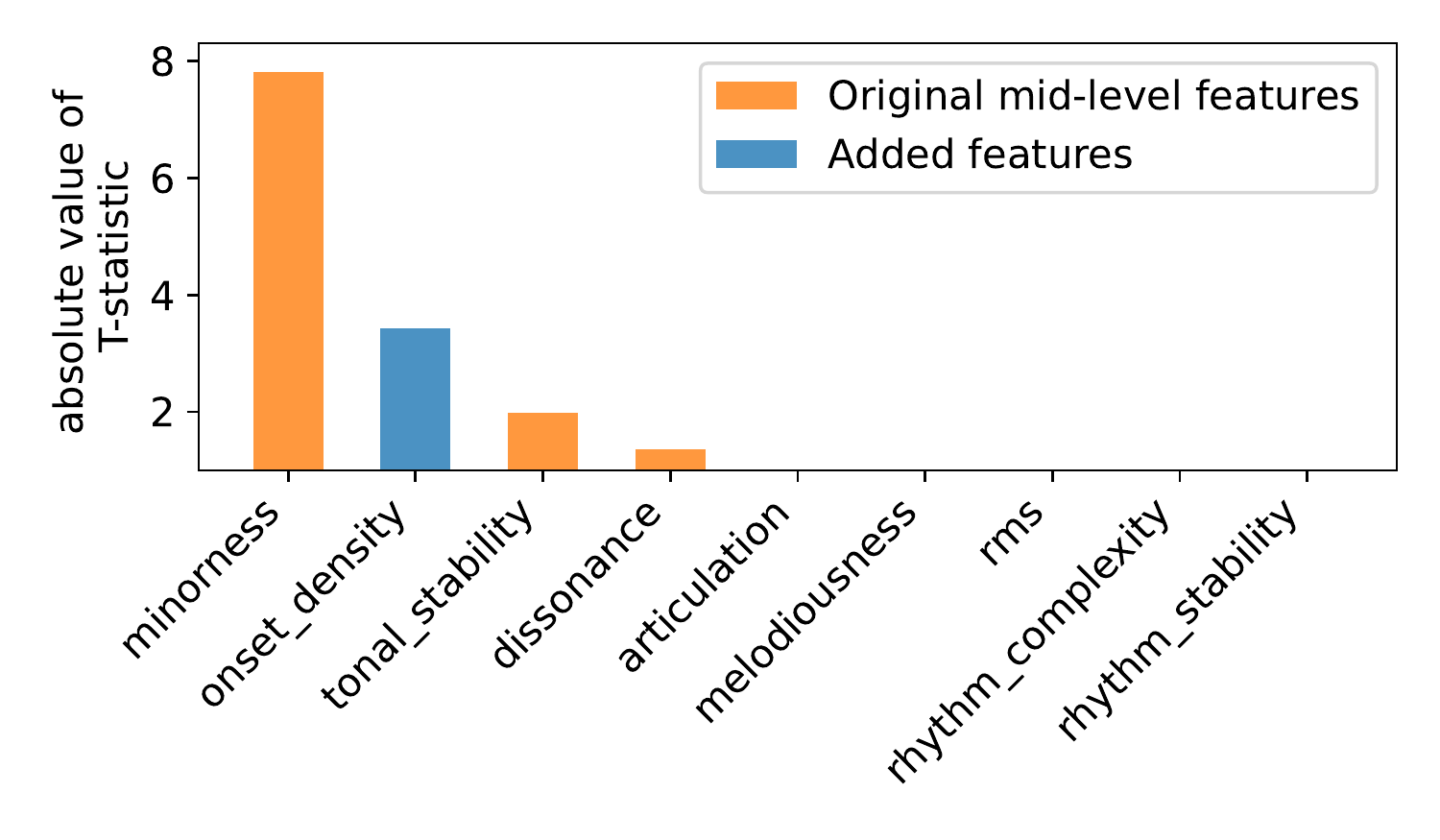}
        \label{fig:tvals_valence}   
    \end{subfigure}
    \vspace{-5mm}
    \caption{Feature importance values for the augmented mid-level feature set, analysed using the T-statistic. For each feature in a linear regression model, the T-statistic is defined as its estimated weight scaled with its standard error.}
    \label{fig:tvals}
\end{figure}

Further, we examine the feature importances for the prediction of arousal and valence using T-statistics (or T-values), plotted in Figure~\ref{fig:tvals}. The T-statistic of a feature in a regression model is defined as its estimated weight scaled with its standard error.

We note that for both arousal and valence, using the combined (9)-mid-level feature set gives the best result. Our findings point to a future direction of work on learning perceived speed and perceived dynamics from actual human ratings as a means to improve music emotion recognition algorithms grounded on human perception.

\begin{table}
\footnotesize
\captionsetup{font=small}
\centering
  \begin{tabular}{l|r|r}
  \toprule
    \multirow{2}{*}{} &
      \multicolumn{2}{c}{Adjusted R$^2$} \\
    
    Feature Set & \text{Arousal} & \text{Valence}  \\
    \midrule
    The (7)-mid-level feature set & 0.68 & 0.63  \\
    Onset density and RMS amplitude & 0.74 & 0.39 \\
    The (9)-mid-level feature set & \textbf{0.79} & \textbf{0.65}  \\
    
    \bottomrule
  \end{tabular}%
  \caption{Performance of the different feature sets on modelling the arousal valence of the Bach WTC Book 1 dataset.}
  \label{tbl:perf}
\vspace{-2mm}

\end{table}

\vspace{-2mm}
\section{The Visualisation}\label{sec:thevis}

In a YouTube video \cite{jacob}, Jacob Collier plays the piece ``Danny Boy" on a piano and modifies it according to different emotions shown to him while he is playing. Using our mid-level feature-based emotion model, we continuously (hop length = 1 second; window length = 5 seconds) predict emotions from the audio extracted from a part of the video (0:00 to 2:45 minutes, dealing with what Jacob calls ``basic" or ``tier-1 emotions") and map them onto Russell's circumplex \cite{russell1980circumplex}\footnote{Note that this is a demonstration of our model, and not a detailed analysis of model performance.}. 

The predicted emotion values are visualised together with the original video. To achieve smooth animation, exponential interpolation is utilised to obtain values between actual predicted values at 1-second intervals. Our visualisation video is accessible at \url{https://youtu.be/1Y8wDZfuaKU}.


\begin{figure}
\captionsetup{font=small}
 \centerline{
 \includegraphics[width=\columnwidth]{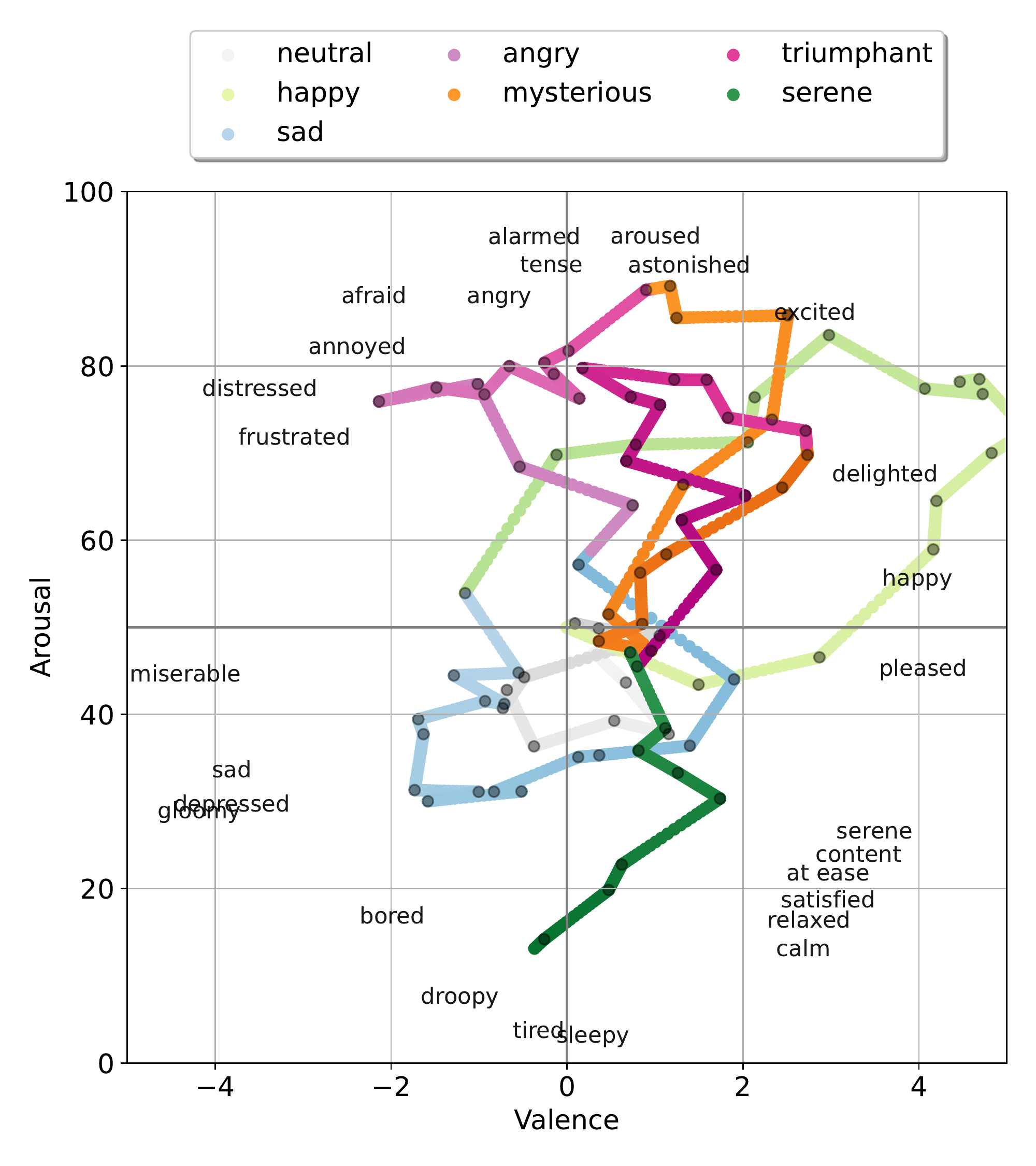}}
 \caption{Full trace of dynamic emotion prediction. Jacob's intended emotions (according to the notated emotion in the original video) are depicted with different colours, and the passage of time is depicted with the shade -- from lightest to darkest.}
 \label{fig:worm}
\end{figure}

\begin{figure}
\captionsetup{font=small}
 \centerline{
 \includegraphics[width=\columnwidth]{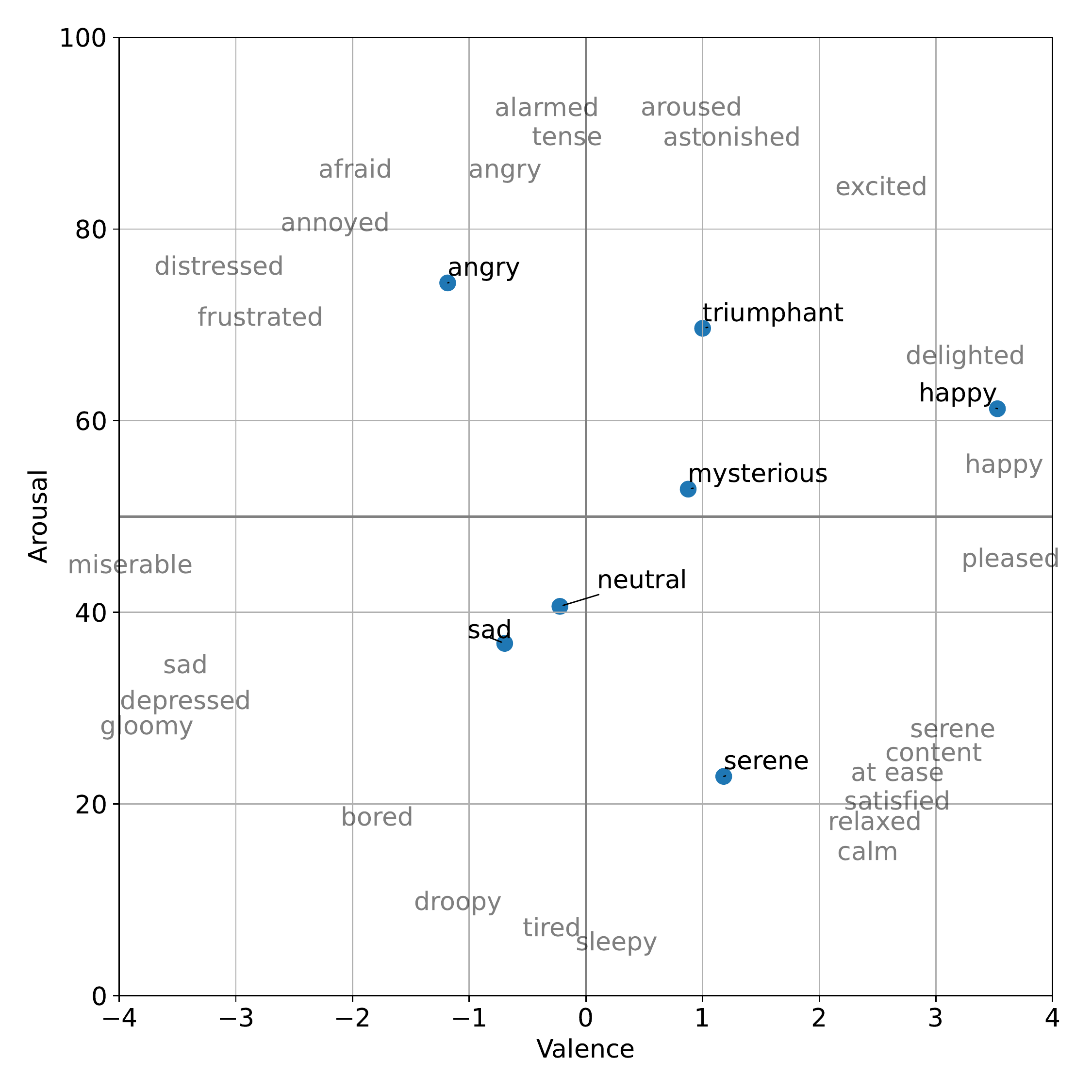}}
 \caption{Static emotion prediction. The predicted emotions are marked
with dark text, and the emotion words of Russell’s circumplex are
marked with light text.}
 \label{fig:static}
\end{figure}

\newpage
\section{Results and Discussion}
\begin{figure}[h!]
\captionsetup{font=small}
     \centering
     \begin{subfigure}[b]{\columnwidth}
         \centering
         \frame{\includegraphics[trim=0 5.1cm 0 5.2cm, clip, width=\textwidth]{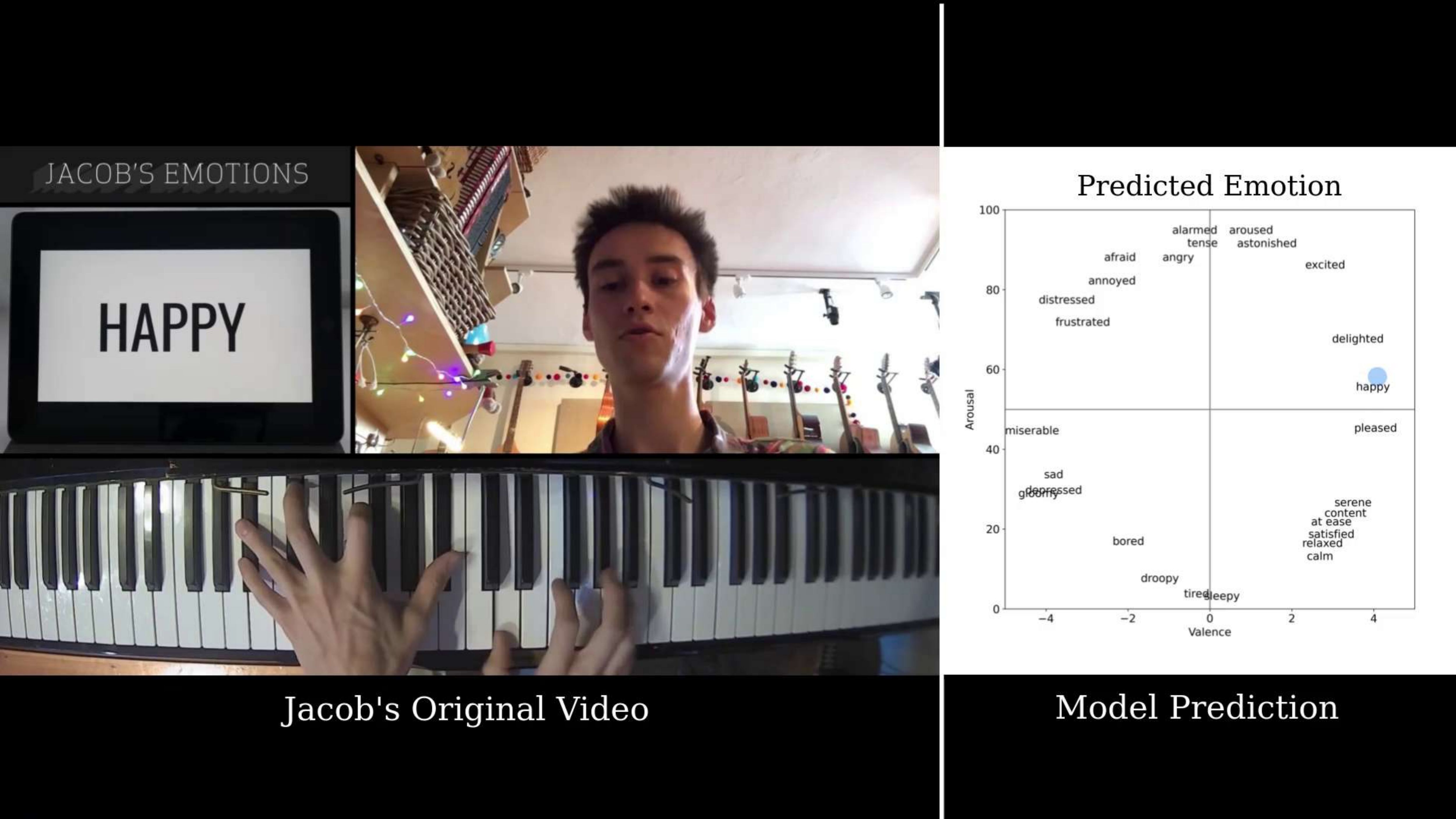}}
		\label{}
     \end{subfigure}\\
     \begin{subfigure}[b]{\columnwidth}
         \centering
         \frame{\includegraphics[trim=0 5.1cm 0 5.2cm, clip, width=\textwidth]{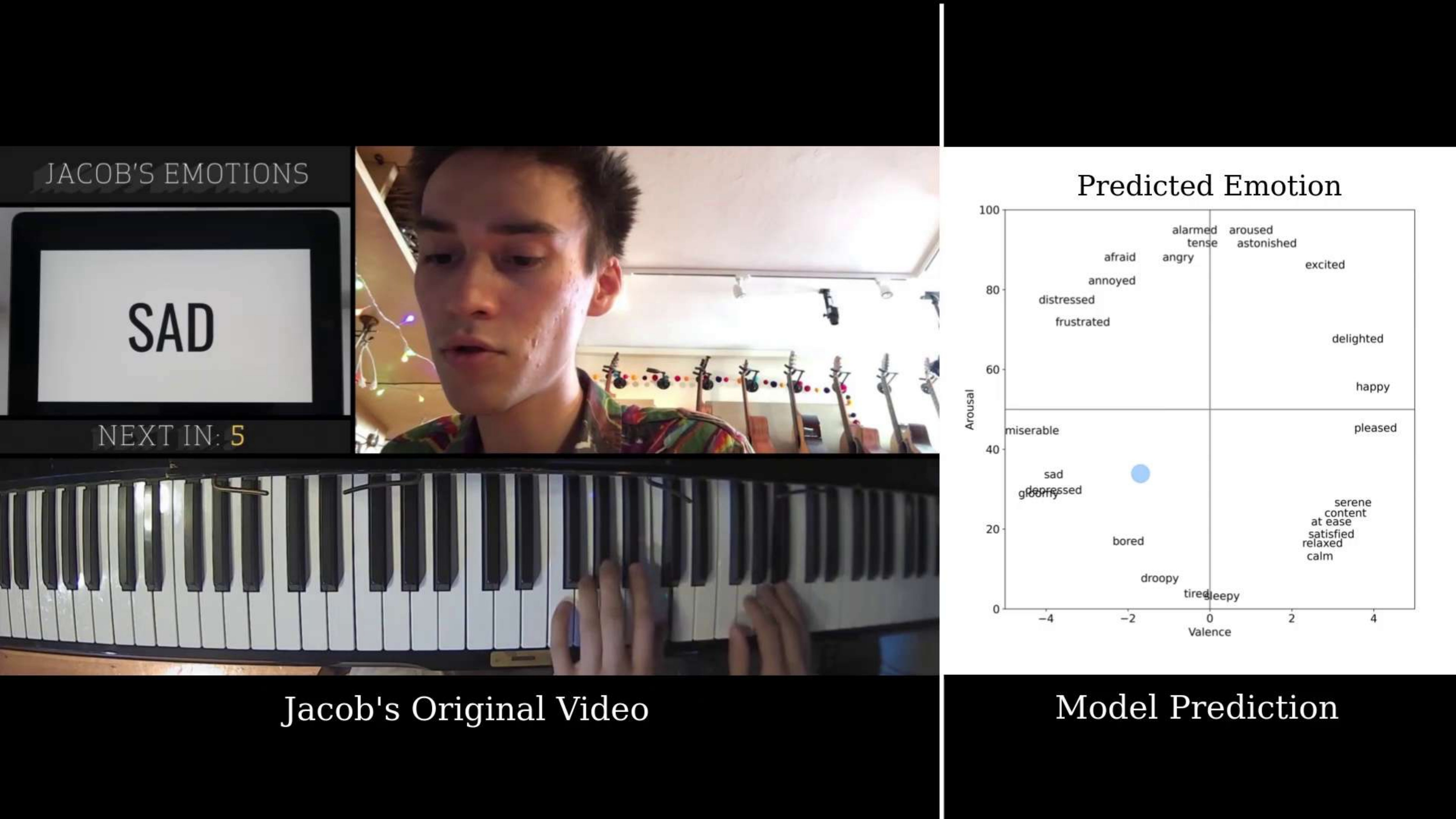}}
		\label{}
     \end{subfigure}\\
     \begin{subfigure}[b]{\columnwidth}
         \centering
         \frame{\includegraphics[trim=0 5.1cm 0 5.2cm, clip, width=\textwidth]{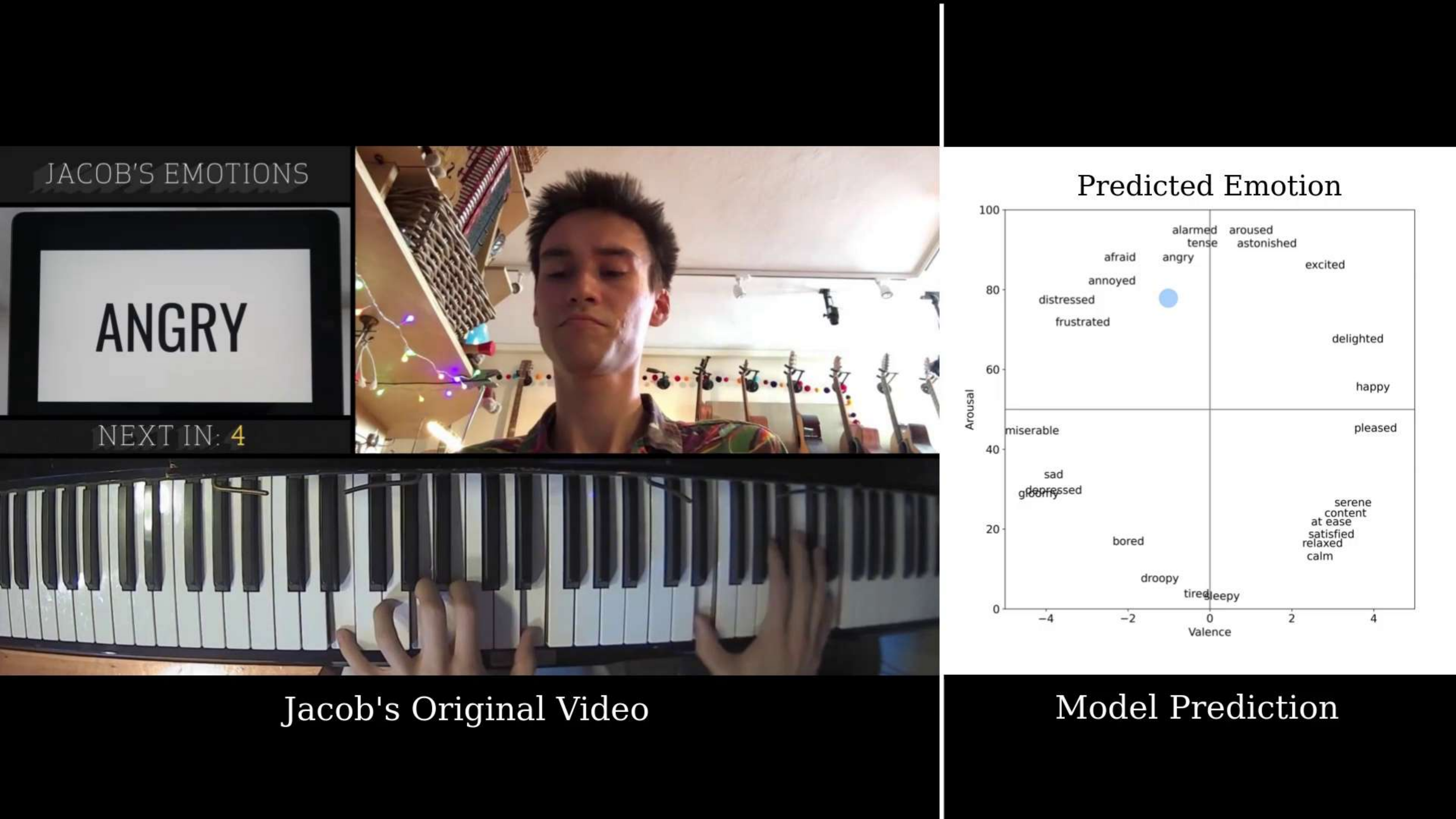}}
		\label{}
     \end{subfigure}\\
     \begin{subfigure}[b]{\columnwidth}
         \centering
         \frame{\includegraphics[trim=0 5.1cm 0 5.2cm, clip, width=\textwidth]{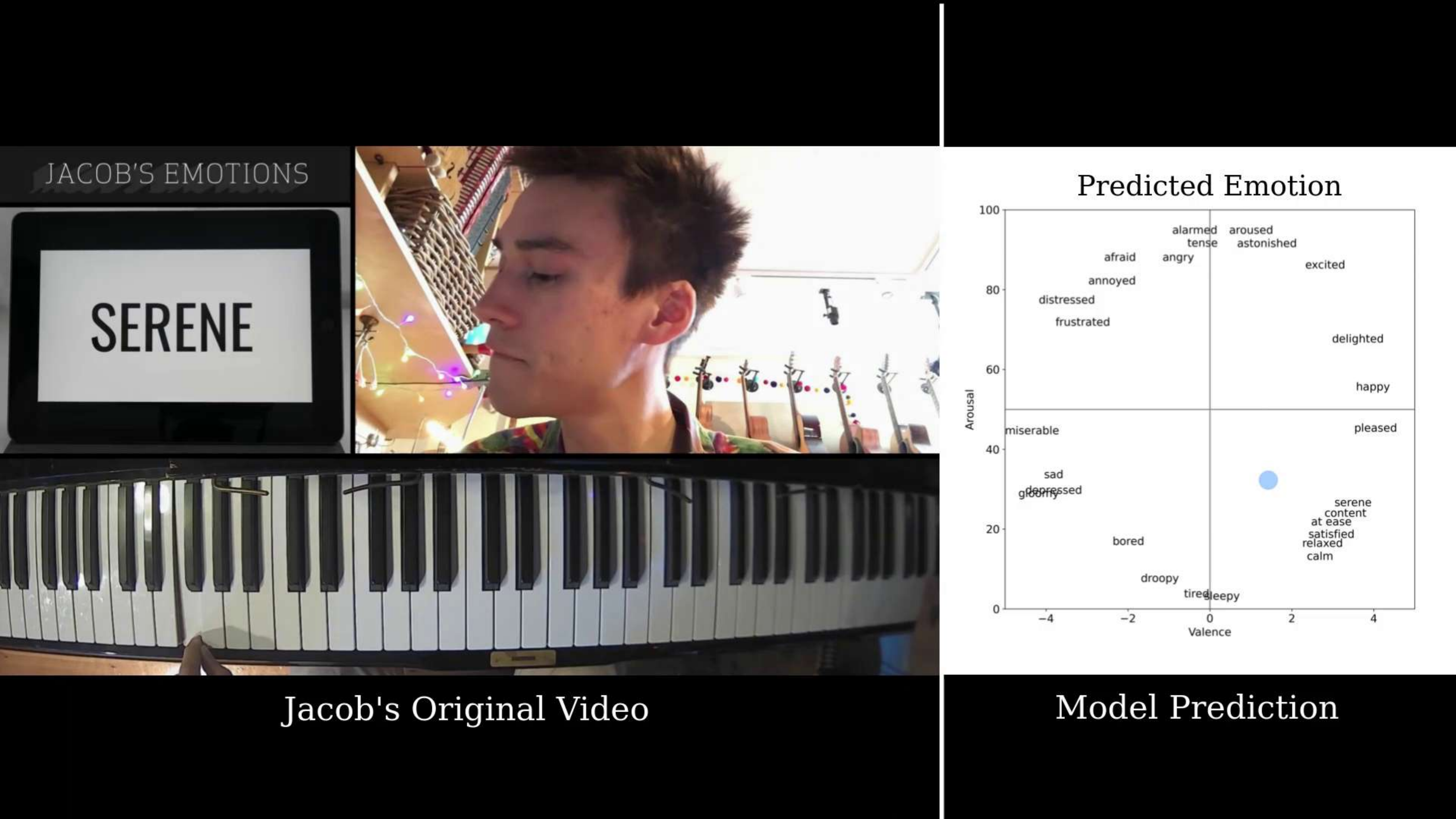}}
		\label{}
     \end{subfigure}\\
        \caption{Screenshots during different times of Jacob Collier's performance video, overlaid with the corresponding predicted emotions.}
        \label{jc_emotions_part1}
\end{figure}

The complete trace of the prediction point is demonstrated in Figure~\ref{fig:worm} while Figure~\ref{jc_emotions_part1} exhibits three frames from the video overlaid with the corresponding predicted emotions. We can see that the predicted emotions match closely with the intended emotions (``Jacob's Emotions"). 

We also obtain static emotions -- the audio sections corresponding to each of the seven emotions are extracted and used as individual audio inputs. In this case, we use our standard input length (15-second) mid-level feature model, with the input audio being looped if it is less than 15 seconds, and the predictions for successive windows with a 5-second hop averaged if it is more than 15 seconds. These static emotion predictions are shown in Figure~\ref{fig:static}, where the predicted points are annotated with the intended emotion for each.

This visualisation experiment provides a proof-of-concept for further investigations into the decoding of intended emotions using computer systems. Interestingly, we observe that a simple linear regression model with only a handful of mid-level features as inputs (7 original plus 2 new), trained on a small dataset of 288 Bach WTC examples, successfully predicts intended emotions for a markedly different set of performances in a satisfactory manner. This points to the robustness of the (9)-mid-level features and of our (7)-mid-level feature model, and to the impressive capacity of these features to reflect encoded music emotion. 

In conclusion, this paper provides promising evidence that mid-level features are effective at capturing emotions conveyed by music performances and encourages further exploration to fully understand the potential of mid-level features as a means of modelling more complex emotions.

\newpage

\section{Acknowledgments}
This research was supported by the European Research Council (ERC) under the European Union's Horizon 2020 research and innovation programme, grant agreements No 670035 (``Con Espressione") and 101019375 (``Whither Music?").

\bibliography{ISMIRtemplate}

\end{document}